\documentclass[reqno]{amsart}

\usepackage{cite}
\usepackage{amsmath,amsfonts,amsthm}
\numberwithin{equation}{section}

\newtheorem{thm}{Theorem}
\newtheorem{lemma}{Lemma}

\newcommand{\Z}{\mathbb Z}
\newcommand{\Hi}{\mathcal H}
\newcommand{\e}{\mathrm e}
\newcommand{\im}{\mathrm i}

\newcommand{\1}{{\mathbf 1}}
\newcommand{\R}{\mathbb R}
\newcommand{\C}{\mathbb C}
\newcommand{\di}{\mathrm d}

\DeclareMathOperator*{\sotlim}{SOT-lim}

\def\norm#1{\left \| #1 \right \|}
\newcounter{list}
\begin{document}
\flushbottom
\title{A strong operator topology adiabatic theorem}
\author{Alexander Elgart and Jeffrey H. Schenker}
\date{February 25, 2002}
\begin{abstract}
We prove an adiabatic theorem for the evolution of spectral data
under a weak additive perturbation in the context of a
system without an intrinsic time scale.  For continuous functions
of the unperturbed Hamiltonian the convergence is in norm while
for a larger class functions, including the spectral projections
associated to embedded eigenvalues, the convergence is in the
strong operator topology.
\end{abstract} \maketitle
\section{Introduction}

The aim of this paper is to give a slightly 
new perspective on adiabatic theorems
related to systems without intrinsic time scales.  We consider 
convergence in the strong operator topology and hope to convince the
reader that this is a natural setting for adiabatic theorems
when there is no intrinsic notion of ``slowness.''

The adiabatic theorem of quantum mechanics describes the behavior of 
a non\-autonomous system driven by means of slowly altered external 
field. An illustrative example is the case of a spin-$1/2$ 
particle (two level system)
coupled to a rotating magnetic field of constant
amplitude. The evolution of this system is 
generated by the time dependent 
Hamiltonian $H(t):= \vec\sigma\cdot\vec B(t)$, where $\{\sigma_i\}$ 
are the Pauli matrices 
and $\vec B$ is the rotating magnetic field. 
There are two time scales here: the 
inverse of the rate at which magnetic field is changing, 
$t_1:=\vert B \vert / \vert\dot B \vert$, 
and the intrinsic time scale of the two level system, $t_2:=1/ |B|$, which
is linked to the gap between the energy levels of the instantaneous Hamiltonian. 
It is natural to say that the system 
changes ``slowly'' if the ratio $t_2/t_1$ is small. Adiabatic 
theory \cite{BF} implies in this context  
that if initially the system is in a 
stationary state -- for instance, with the spin parallel 
to the magnetic field $\vec B(0)$ -- then it will stay close to an
instantaneous stationary state -- i.e., parallel to the direction of 
$\vec B(t)$.  
``Close'' here means that the transition amplitude to 
the second stationary state -- anti-parallel to $\vec B(t)$ -- 
is bounded from above by a function of 
the ratio $t_2/t_1$ which vanishes at zero.

In general, the subject of quantum adiabatic theory is 
the unitary evolution which solves 
an initial value problem (Schr\"odinger equation) 
of the form
\begin{equation}\label{eq:IVP}
\begin{cases}
\im \dot U_\tau(t) \ =& \ H  ( t / \tau) U_\tau(t)  
\; , \quad t \in [0, \tau]  \\
  U_\tau(0) \ =& \ \1
\end{cases}  \; ,
\end{equation}
with a time dependent self adjoint operator $H(s)$ for $s \in [0,1]$. 
The parameter $\tau$ is supposed to provide a scale to measure 
the ``slowness'' of the system, and adiabatic theory is concerned with  
the limit  $\tau \rightarrow \infty$. Strictly speaking, 
to determine what is meant by ``slow,'' we need a second time scale 
coming from the structure of the 
system -- e.g., a spectral gap as in the above example.

When applicable, the adiabatic theorem
states that
\begin{equation}\label{lim}
\lim_{\tau \rightarrow \infty}U_\tau(\tau s) f(H(0)) 
U^\dagger_\tau(\tau s)  \ = \ f(H( s)) 
\; , \quad s \in [0, 1]   \; .
\end{equation} 
However, to be precise 
we should indicate in what topology this limit is taken, 
and this issue is the heart of this work.
 
As far as we know, to date eq.~\eqref{lim} has
always been understood in the norm sense. This choice 
has been well justified since for the systems in question the
the meaning of slowness was intrinsic. 
For instance, if the function $f$ in \eqref{lim} is a projection 
to a spectral band separated by a finite gap from the rest of the 
spectrum, then the adiabatic theorem holds in the norm sense \cite{ASY,N}. 
In this case, the inverse of the spectral gap 
provides an intrinsic time scale. 

There are also examples of systems without a spectral gap but nonetheless 
a clearly defined intrinsic time scale. The first example is the
so called level crossing situation: Imagine that two non-degenerate 
eigenvalues of the instantaneous Hamiltonian cross each other at 
some time. Although the spectral gap vanishes at the crossing, there 
is an intrinsic time scale coming from the relative slope of the eigenvalues. 
There is an adiabatic theorem in this case which holds
in the norm sense \cite{hag}.  Recently this was extended
to systems with an infinite number of crossings in finite time 
\cite{joye}.  Another example is a system with
dense point spectrum perturbed by a finite 
rank operator, considered in ref. \cite{ahs}.  
There, the time scale is related not to the gap between energies (which
may be arbitrarily small) but the gap multiplied by
the overlap between the corresponding eigenstates 
coupled through the perturbation.  
Our final example is a system with an eigenvalue of
finite degeneracy at the threshold of, say, continuous spectrum as
considered in \cite{b,ae1}.  Here an intrinsic time scale can be 
extracted from the H\"older continuity of the {\em continuous} part of 
the spectral measure in the vicinity of the eigenvalue. 
In all situations above, the 
adiabatic theorem holds in the norm topology, and, more or less, these
examples exhaust the known results on the subject.\footnote{
There is an extensive literature on adiabatic theory, much of which 
is not cited here.  Several
extensive reviews have appeared recently
(see \cite{ae1,JP}).}

The present paper is concerned with the adiabatic theorem for a system without
an intrinsic time scale.  We are motivated
by problems encountered in the analysis of the Quantum Hall Effect (QHE) in
which one considers a time dependent perturbation of a system with dense
point spectrum. Unlike in ref.~\cite{ahs}, the perturbation is {\em not} finite rank
which has the consequence, as was pointed out to us by Michael Aizenman, 
that one does not expect the adiabatic theorem to be true in the norm operator
topology in that context. A somewhat 
simpler example of the phenomenon which occurs there is provided by 
a direct sum of infinitely many non-interacting
systems each of which with its own characteristic time scale.  Once the
adiabatic parameter $\tau$ is larger than the time scale of an individual
system, that subsystem is close to the adiabatic limit.  However, if the
sequence of time-scales is unbounded, there is no notion of 
slowness which holds for the {\em whole} system. We discuss this example in
more  detail in Section \ref{sec:example}.

We consider in this paper a family of Hamiltonians of the form
\begin{equation}\label{eq:SE} 
H_{\tau}(t/\tau) \ = \ H_{o}\ + \ \frac{1}{\tau}
  \Lambda(t/\tau) 
\end{equation}
where $H_o$ and $\Lambda(s)$, for $s \in [0,1]$, are self adjoint
operators. The particular form for the time dependence is formulated with the
QHE in mind. 

The usual adiabatic framework 
involves a Hamiltonian which depends on $\tau$ only through the
rescaling of time -- see eq.~\eqref{eq:IVP}.  
The evolution consider here is equivalent, via a unitary transformation,
to the solution of \eqref{eq:IVP} with 
$H(s) = V(s) H_{o} V^{\dag}(s)$, where $\Lambda(s)$ is the
generator of $V(s)$: 
\begin{equation}
\im \dot V(s) \ = \ \Lambda(s) V(s) \; , \quad V(0) \ = \ \1 \; .
\end{equation}
In physical literature, this description of the dynamics
is referred to as the ``interaction picture,'' and has proved useful
in many situations.

We discuss here the limit $\tau \rightarrow \infty$ of a
solution, $A_\tau(t) = U_\tau(t) A(0) U^\dag_\tau(t)$, to the
associated Heisenberg equation
\begin{equation}
\im \dot A_\tau(t) \ = \ \left [H_{\tau}(t/\tau) \, 
, \, A_\tau(t) \right ] \;
\end{equation}
when the initial observable is a function of $H_o$,{\it i.e.},
$A(0) = f(H_o)$. Our main result, Theorem~\ref{at}, states that
\begin{equation}\label{eq:limit}
U_\tau(\tau) f(H_o) U^\dag_\tau(\tau) \ \longrightarrow \ f(H_o)
\end{equation}
for a wide class of functions $f$.

The topology in which eq.~\ref{eq:limit} holds depends on the
continuity of $f$ relative to the spectral properties of $H_o$:
for continuous functions we obtain norm convergence while for a class
of discontinuous functions we obtain strong operator convergence. Let
us recall that a family $\tau \mapsto A_\tau$ of operators
converges to $A$ in the {\it strong operator topology} (SOT) if
\begin{equation}
\lim_{\tau \rightarrow \infty} A_\tau \, \psi \  = \ A \, \psi
\end{equation}
for every $\psi \in \Hi$ and converges in norm if
\begin{equation}
\lim_{\tau \rightarrow \infty} \norm{A_\tau - A} \ = \ 0 \; .
\end{equation}
We denote SOT convergence by ``$\sotlim A_\tau = A $''.

The remainder of this paper is organized as follows.  
Sections~\ref{sec:thm}, \ref{sec:proof} and \ref{sec:BV} are devoted to the
statement and proof of our main result, Theorem~\ref{at} in
section~\ref{sec:thm}. 
In section~\ref{sec:example} we describe an example which shows that 
the norm topology is inadequate when we
consider discontinuous functions of $H_o$. Finally, in
section~\ref{sec:conjecture} we present a stronger result which
holds when $H_o$ has pure point spectrum.

\section{The theorem and all we can show with the
resolvent}\label{sec:thm}

Before we state Theorem~\ref{at}, let us recall the definition of
certain classes of functions $f:\R \rightarrow \C$:
\begin{list}{(\arabic{list})} {\usecounter{list}
\setlength{\leftmargin=24pt} \setlength{\labelsep=12pt}
\setlength{\itemindent=12pt}} \item Let $C_b$ denote the bounded
continuous functions.
\item Let $C_o$ denote those functions in $C_b$ which vanish at
$\pm \infty$.
\item Let $BV$ denote the functions of {\em bounded variation},
i.e., functions $f$ for which
\begin{equation} {\rm
Var}(f) \ := \ \sup_{n \ge 1} \ \sup_{x_o < \cdots < x_n \in \R} \
\sum_{j=1}^n |f(x_j) - f(x_{j-1})|   \ < \ \infty \; .
\end{equation}
A function in $BV$ can have only countably many points of
discontinuity. We direct the reader to \cite[Ch. 3]{folland} 
for a detailed discussion of $BV$.
\end{list}

\begin{thm}\label{at} Let $H_o$ be a self adjoint operator
and suppose that the time evolution $U_\tau$ satisfies the initial
value problem \eqref{eq:IVP} with $H_{\tau}(t/\tau) = H_{0} + (1/\tau) 
\,
\Lambda(t/\tau)$ where $\Lambda(\cdot)$ is a self adjoint family which
is $L^1$ in norm: $\int_0^1 \di s \norm{\Lambda(s)} < \infty$.
Given a
measurable function $f$, consider the statement
\begin{equation}\label{eq:thmlimit}
\lim_{\tau \rightarrow \infty} W_\tau(s)f(H_{0})W_\tau^\dag(s) \ =
\ f(H_o) \; , \mbox{ uniformly for }  s \in [0,1] \; ,
\end{equation}
where $W_\tau$ is the evolution at scaled time, $W_\tau(s) =
U_\tau(\tau \cdot s)$ .
\begin{list}{(\arabic{list})} {\usecounter{list}
\setlength{\leftmargin=24pt} \setlength{\labelsep=6pt}
\setlength{\itemindent=-6pt}}
\item If $f \in C_o$ then eq.~\eqref{eq:thmlimit} is true
in the operator norm topology .
\item If $f = g + h$ with $g \in C_b$ and $h \in BV$ then
eq.~\eqref{eq:thmlimit} is true in the strong operator topology.
\end{list}
\end{thm}
\noindent{\it Remarks}:
\begin{list}{(\arabic{list})}
{\usecounter{list}  \setlength{\leftmargin=24pt}
\setlength{\labelsep=12pt} \setlength{\itemindent=12pt}}

\item Operators $A_\tau(s)$ are said to converge uniformly to
$A$ in the strong operator topology if 
\begin{equation} \lim_{\tau \rightarrow \infty}
\sup_{s} \norm{A_\tau(s) \psi - A \psi} \ = \ 0
\end{equation}
for every $\psi \in \Hi$.  Uniform norm convergence is defined
similarly.

\item Among the usual operator topologies, i.e., the norm topology 
as well as the strong and weak operator topologies, 
the strong operator topology is the strongest in
which we can expect an adiabatic limit for discontinuous functions
of $H_o$. In section~\ref{sec:example} we describe an elementary
example of a system for which $W_\tau(s) f(H_o) W_\tau^\dag( s)$
fails to converge in the norm topology.

\item If the operator $H_o$ is unbounded, the distinction between
$C_o$ and $C_b$ is meaningful.  Functions in
$C_b$ may be ``discontinuous at infinity'' which explains the loss
of norm convergence.

\item Among the functions of bounded variation are the {\em
Kr\"onecker} delta functions: $\delta_E(x) = 1$ if $x = E$ and $0$
otherwise. Thus we obtain an adiabatic evolution for the spectral
projection associated to any eigenvalue -- even if it has infinite
degeneracy and is embedded in the essential spectrum!

\item As described above, 
the standard adiabatic theorems describe the limiting behavior
of the Schr\"odinger evolution for a system having a gap in its
spectrum with initial data being a spectral projection onto an
energy band.
A projection onto a spectral band is a {\em continuous} function
of $H_o$, thus the convergence occurs in the norm topology. In
such a setting it is possible to find an explicit bound on the
rate of convergence in eq.~\ref{eq:thmlimit} (see, for example,
eq.~\eqref{eq:explicit} and Lemma~\ref{O}).

\item 
Schr\"odinger equations with a Hamiltonian of the form considered
here find direct
application in the description of the motion of a quantum particle
in a time dependent potential energy.   In that case, $H_o$
describes the motion of the particle in the absence of time
dependent terms and is generally the Laplacian or some
perturbation thereof, possibly discretized, the underlying Hilbert
space being $\ell^2(\Z^d)$ or $L^2(\R^d)$. The time dependent term
$\Lambda(t)$ is the operator of multiplication by a bounded
function $\Lambda(x,t)$. Theorem~\ref{at} is relevant
to the adiabatic evolution of an ensemble of non-interacting
particles with Fermi statistics. The observables, in this case,
are the Fermi-Dirac distributions $F_{\mu,\beta}(H_o) =
\frac{1}{1 + e^{\beta(H_o - \mu)}}$ at positive temperatures and
the spectral projections $\chi(H_{o} \le \mu)$ and/or 
$\chi(H_{o} < \mu)$ at zero temperature.  We obtain an adiabatic 
evolution even if there is an eigenvalue at the chemical potential $\mu$!
\end{list}

The heart of the matter lies in the proof of Theorem~\ref{at}
under the additional assumption that $\Lambda$ is boundedly
differentiable in norm, i.e., that
\begin{equation}
  \dot \Lambda(s) \ := \  \lim_{h \rightarrow 0} 
  \frac{\Lambda(s+h) - \Lambda(s)}{h} 
\end{equation}
exists in the norm topology for each $s \in (0,1)$ and $\sup_s 
\norm{\dot \Lambda(s)}$ is finite.  The extension  
to general $\Lambda$ is accomplished by a standard mollifier argument.

Specifically, we choose a positive smooth function $\phi(s)$ 
with compact support such that $\int \phi = 1$ and set
$\phi_\epsilon(s) = \epsilon^{-1} \phi(s/\epsilon)$.  Then
\begin{equation}
  \Lambda_{\epsilon}(s) \ = \ \int \di s' \phi_\epsilon(s - s')  
  \Lambda(s') \; ,  
\end{equation}
is boundedly differentiable and 
\begin{equation}
\lim_{\epsilon \rightarrow 0} \int_0^1 \norm{\Lambda_\epsilon(s) - \Lambda(s)}
\ = \ 0 \; .
\end{equation}
Let $U_{\tau, \epsilon}$ be the solution to the 
IVP \eqref{eq:IVP} with 
$H_\tau(t/\tau) = H_o + (1/\tau)\Lambda_\epsilon(t/\tau)$ and
set $W_{\tau,\epsilon}(s) = U_\tau(\tau \cdot s)$. Then,
\begin{equation}
  \frac{\di}{\di s} W_{\tau,\epsilon}^\dag(s) 
  W_\tau(s) \ = \ 
   W_{\tau,\epsilon}^\dag(s) \, 
  \left ( 
    \Lambda_\epsilon(s) - \Lambda(s) 
  \right ) \,
  W_\tau(s) \; .
\end{equation}
>From this it follows that
\begin{equation}
  \norm{W_{\tau,\epsilon}(s) - W_\tau(s)}
  \ \le \ \int_0^1 \di s \norm{ \Lambda_\epsilon(s) - \Lambda(s)} \ 
  \rightarrow \ 0 \; ,
\end{equation}
i.e., $W_{\tau,\epsilon}(s)$ converges to $W_\tau(s)$ 
uniformly in {\em $s$ and $\tau$}.  By a standard ``2-$\epsilon$'' argument the 
theorem now follows for $\Lambda$ in $L^1$ once it is verified
for $\Lambda_\epsilon$.  Hence, it suffices to show Theorem~\ref{at} 
for differentiable $\Lambda$.

\vspace{1em}

{\em Throughout the rest of the paper, $\Lambda$ will denote a uniformly bounded 
  self-adjoint family which is differentiable in the norm topology with
  a uniformly bounded derivative $\dot \Lambda$.}

\vspace{1em}

The remainder of this section is devoted to the proof of those parts of
Theorem~\ref{at} which follow from norm resolvent convergence. This
part of the proof is very elementary but is also unrelated to 
the arguments in the subsequent sections.  In section~\ref{sec:proof}, we
present a Lemma~\ref{lem:ibp}, which states that the portion of
Theorem~\ref{at} related to functions of bounded variation
($BV$) may be reduced to a statement about spectral projections. A
proof of this Lemma, based on ideas that go back to
Kato~\cite{k}, is also given in section~\ref{sec:proof}. In
section~\ref{sec:BV} we prove Lemma~\ref{lem:ibp}.

For a great many functions, $f$, the conclusion of Theorem~\ref{at} 
-- i.e., eq.~\eqref{eq:thmlimit} --  follows from
well known convergence theorems and a simple formula --
eq.~\eqref{eq:resolvent} -- which shows that
\begin{equation}\label{eq:nrc}
\sup_{s \in [0,1]} \norm{W_\tau(s) (H_o - z)^{-1} W_\tau^\dag( s)
- (H_o - z)^{-1} } \ \longrightarrow \ 0
\end{equation}
for every $z \not \in \R$, which is to say that $W_\tau(s) H_o
W_\tau^\dag(s) \rightarrow H_o$ uniformly in $s$ in the ``norm
resolvent sense''. The implications of norm resolvent convergence
for Theorem~\ref{at} are that
\begin{list}{(\arabic{list})}
{\usecounter{list}  \setlength{\leftmargin=24pt}
\setlength{\labelsep=12pt} \setlength{\itemindent=12pt}}
\item Eq.~\ref{eq:thmlimit} holds in the norm topology for $f \in
C_o$ \cite[Thm. VIII.20] {reed&simon}.
\item  Eq.~\ref{eq:thmlimit} holds in the strong operator topology for
$f \in C_b$ or when $f$ is the characteristic function of an open
interval $(a,b)$ provided that $a$ and $b$ are not eigenvalues of
$H_o$.  This follows from \cite[Thm.
VIII.20 and VIII.24] {reed&simon} since ``strong resolvent
convergence'' is implied by ``norm resolvent convergence.''
\end{list}
What is remarkable is that with some additional work we can prove
eq.~\ref{eq:thmlimit}, for example, when $f$ is
the characteristic function of an open interval $(a,b)$ and
{\em one or both of $a,b$ is an eigenvalue with arbitrary
degeneracy}.

To verify eq.~\eqref{eq:nrc}, we use the identity
\begin{multline}\label{eq:resolvent}
\left ( H_\tau(s)  - z \right )^{-1} \  - \ W_\tau(s) \left (
H_\tau(0) - z \right )^{-1} W_\tau(s)^\dag
\\ =  \ W_\tau(s)  \int_0^{s}  W_\tau(t)^\dag  \left ( \frac{\di}{\di t}
\left ( H_\tau(t) - z \right )^{-1} \right ) W_\tau(t) \, \di t \,
W_\tau(s)^\dag \; ,
\end{multline}
where $H_\tau(s) = H_o + \frac{1}{\tau} \Lambda(s)$.
Eq.~\eqref{eq:resolvent} follows from the fundamental theorem of
calculus and the observation that
\begin{equation}
\frac{\di}{\di t} \Big ( W_\tau(t)^\dag  \left ( H_\tau(t) - z
\right )^{-1} W_\tau(t) \Big ) \ = \ W_\tau(t)^\dag  \left (
\frac{\di}{\di t} \left ( H_\tau(t) - z \right )^{-1} \right )
W_\tau(t) \; .
\end{equation}
Now,  eq.~\eqref{eq:nrc} follows from eq.~\eqref{eq:resolvent}
because the latter implies that
\begin{equation}\label{eq:explicit}
\norm{W_\tau(s) (H_o - z)^{-1} W_\tau(s)^\dag - (H_o - z)^{-1} } \ \le
\ \frac{C}{(\mathrm{Im} z)^2 } \frac{1}{\tau}  \; ,
\end{equation}
since
\begin{equation}
\frac{\di}{\di t} \left ( H_\tau(t) - z \right )^{-1}  \ = \
\frac{1}{\tau} \left ( H_\tau(t) - z \right )^{-1} \,  \dot
\Lambda(t) \, \left ( H_\tau(t) - z \right )^{-1} \; ,
\end{equation}
and
\begin{equation}
\left ( H_\tau(s) - z \right )^{-1} \ = \ \left ( H_o - z \right
)^{-1} - \frac{1}{\tau} \left ( H_\tau(s) - z \right )^{-1}
\Lambda(s) \left ( H_o
 - z \right )^{-1} \; .
\end{equation}

Before we proceed, let us describe an example which demonstrates
that we cannot hope to prove eq.~\eqref{eq:thmlimit} for general $f$
in $BV$ using {\em only} the fact that $W_\tau H_o W_\tau^\dag$ 
converges to $H_o$ in the norm resolvent sense. For this purpose it
is sufficient to produce a sequence of unitary operators $V_n$ such
that $V_n H_o V_n^\dag$ converges to $H_o$ but nonetheless
\begin{equation}
  \sotlim_{n \rightarrow \infty} V_n f(H_o) V_n^\dag \ \neq \ f(H_o)
\end{equation}
for some function $f \in BV$. 

For this purpose,
consider the self adjoint operator on $\ell^2(\Z)$
given in Dirac notation by $H_o = \sum_{m \neq 0} \frac{1}{m} |m
\left > \right < m|$, and  for each $n$ let $V_n$ be the unitary on
$\ell^2(\Z)$ which ``swaps $0$ and $n$'', {\it i.e.},
\begin{equation}
(V_n \psi)(m) \ = \ \begin{cases} \psi(m) & \text{if $m \neq 0,
n$} \\ \psi(0) & \text{if $m = n$} \\ \psi(n) & \text{if $m = 0$}
\end{cases} \; .
\end{equation}
Then $ V_n H_o V_n^\dag = H_o + \frac{1}{n} \left ( |0 \left
> \right < 0| - |n \left > \right < n| \right )$.  So $V_nH_o
V_n^\dag \rightarrow H_o$ in norm, and thus in norm resolvent
sense. Yet,
\begin{equation}
V_n |0 \left > \right < 0 | V_n^\dag \ = \ |n \left
> \right < n| \ \stackrel{ \text{SOT}}{\longrightarrow} \ 0
\quad n \rightarrow \infty \; ,
\end{equation}
and $|0 \left> \right < 0| = P_{0} $, 
the spectral projection of $H_o$ associated to eigenvalue $0$.

\section{SOT convergence for spectral
projections}\label{sec:proof} The claim that
eq.~\eqref{eq:thmlimit} holds whenever $f \in BV$ is, at heart, a
statement about spectral projections as is indicated by the
following lemma:
\begin{lemma}\label{lem:ibp}
Eq.~\eqref{eq:thmlimit} holds in the SOT for every $f \in BV$ if
and only if it holds for all $f$ of the form
$f(x) = \chi(x \le E)$ or $f(x) = \chi(x \ge E)$ with any 
$E \in \R$.
\end{lemma}
\noindent We postpone the proof of Lemma~\ref{lem:ibp} to
section~\ref{sec:BV} and focus here on proving
eq.~\ref{eq:thmlimit} with $f(x) = \chi(x \ge E)$ and $f(x) =
\chi(x \le E)$ for every $E$ in $\R$.

In what follows we fix $E$ and take $P = \chi(H_o \le E)$. The
other case, $\chi(H_o \ge E)$, is handled in exactly the same
way by changing $\le$ to $\ge$ in the appropriate places.  We must
show that for any $\psi \in \Hi$
\begin{equation}\label{eq:adfermproj}
\lim_{\tau \rightarrow \infty} \, \sup_{s \in [0,1]} \norm{ \left
( W_\tau(s) \, P \, W_\tau(s)^\dag \ - \ P  \right ) \psi} \ = \ 0
\; .
\end{equation}

Our argument is stated most readily with the propagator
$W_\tau(t,s) = W_\tau(t) W_\tau^\dag(s)$ -- note that $W_\tau(s) P
W_\tau^\dag(s) = W_\tau(s, 0) P W_\tau(0,s)$ and $P = W_\tau(s,s)
P W_\tau(s,s) $. We would like to compare $W_\tau(s,t)$ with the
propagator associated to $H_o$, so we define
\begin{equation}
\Omega_\tau(t,s) \ := \ \e^{\im \tau (t-s) H_o} W_{\tau}(t)
W_\tau(s)^\dag \; .
\end{equation}
Since $\Omega_\tau(t,s)$ is unitary and 
the exponential of $H_o$ commutes with $P$ and
\begin{equation}
\begin{split}
\norm{\left ( W_\tau(s) P W_\tau(s)^\dag - P \right ) \psi} \ =& \
\norm{\left ( \Omega_\tau(0,s)^\dag P \Omega_\tau(0,s) \ - \ P
\right ) \psi } \\  =& \ \| [P, \Omega_\tau(0,s)]\psi \| \; .
\end{split}
\end{equation}
Finally, because $P$ is a projection
\begin{equation}\label{eq:ppbar}
[P, \Omega_\tau(t,s)] \ = \ P \, \Omega_\tau(t,s) \bar P - \bar P
\, \Omega_\tau(t,s) P \; ,
\end{equation}
where $\bar P = \1 - P$. Therefore, eq.~\ref{eq:adfermproj} will
follow if we can verify that both terms on the right side of
eq.~\eqref{eq:ppbar} uniformly converge to zero in the SOT.

Consider the first term.  Let $P_\Delta: =\chi(E < H_{0} < E +
\Delta)$, then
\begin{equation}
P \, \Omega_\tau(t,s) \bar P \ = \ P \,  \Omega_\tau(t,s) ( \bar P
- P_{\Delta}) \ + \ P \,\Omega_\tau(t,s) P_{\Delta}.
\end{equation}
We will see below (Lemma~\ref{O}) that the operator norm of $P
\Omega (P - P_\Delta)$ is uniformly bounded by $1/\tau \Delta$.
Thus given $\psi \in \Hi$
\begin{equation}
\norm{P \, \Omega_\tau(t,s) \bar P \psi} \ \le \ \frac{C}{\tau
\Delta} \norm{\psi} \ + \ \norm{P_\Delta \psi} \; .
\end{equation}
If, for instance, $\Delta=1/\sqrt\tau$ then both terms converges
to zero since $\sotlim P_\Delta = 0$ -- whether or not there is an
eigenvalue at $E$.

The second term of \eqref{eq:ppbar} requires a little more care.
Because $E$ may be an eigenvalue, we need to isolate the
contribution from the associated projection $P_E = \chi(H_o = E)$.
Let $P_\Delta' = \chi(E -\Delta < H_o < E)$ and consider
\begin{equation}
\bar P \, \Omega_\tau(t,s) P \ = \ \bar P \, \Omega_\tau(t,s) ( P
- P_{\Delta}' - P_E)  \ + \ \bar P \, \Omega_\tau(t,s) P_{\Delta}'
\ + \ \bar P \, \Omega_\tau(t,s) P_E \; .
\end{equation}
As above, if we take $\Delta = 1/\sqrt \tau$ then the first and
second terms tend uniformly to zero. That the third term also
converges to zero is the content of the following lemma:
\begin{lemma}\label{lem:PE}
Let $P_E := \chi(H_o = E)$.  Then $ (\1 - P_E ) \,
\Omega_\tau(t,s) P_E $ uniformly tends  to zero in the strong 
operator topology.
\end{lemma}
\begin{proof}
The operator $\Omega_\tau(t,s)$ satisfies a Volterra equation
\begin{equation}\label{eq:volterra}
\Omega_\tau(t,s)\ = \1 + \int_s^t \di r \ K_\tau(r,s)
\Omega_\tau(r,s) \; ,
\end{equation}
with
\begin{equation}
  K_\tau(r,s) \ = \ - \im \e^{\im\tau(r-s)H_o} \Lambda(r) \e^{\im
  \tau(s-r)H_o}  \; .
\end{equation}
By iterating eq.~\eqref{eq:volterra} we obtain a norm convergent
series
\begin{equation}\label{eq:series}
\Omega_\tau(t,s) \ = \ \sum_{n = 0}^\infty A_\tau^{n}(t,s) \;
\end{equation}
where
\begin{equation}\label{eq:intforA}
A_\tau^n(t,s) \ = \ \idotsint\displaylimits_{s \le r_n \le \ldots
\le r_1 \le t} \di r_1 \ldots \di r_n K_\tau(r_1,s) \ldots
K_\tau(r_n,s)
 \; .
\end{equation}
Since $A_\tau^n(t,s)$ is obtained by integrating a product of $n$
factors of $K$ over a simplex of volume $(t-s)^n / n!$ we have the
elementary norm bound
\begin{equation}\label{eq:normbound}
\norm{A_\tau^n(t,s)} \ \le \ \frac{1}{n!} \kappa^n (t-s)^n \; ,
\end{equation}
where $\kappa = \sup_r \norm{\Lambda(r)}$.

Using dominated convergence, 
we see from \eqref{eq:series}, \eqref{eq:normbound},
that it suffices to show for each $n$ that $\bar P_E
A_\tau^n(t,s) P_E \rightarrow 0$ uniformly in the SOT. This may be
proved as follows. First note that
\begin{align}
P_E K_\tau(r,s) P_E \ =& \ - \im P_E \Lambda(r) P_E \; , \\ \bar
P_E K_\tau(r,s) P_E \ =& \ - \im \bar P_E \e^{\im \tau (r-s)(H_o-
E)} \Lambda(r) P_E \; .
\end{align}
Next observe that
\begin{equation}
\int_s^r \di r' \bar P_E \e^{\im \tau (r'-s)(H_o- E)} \
\longrightarrow \ 0
\end{equation}
uniformly in the strong operator topology 
from which it follows via integration by parts that
\begin{equation}\label{eq:barPEB}
\int_s^r \di r' \bar P_E \e^{\im \tau (r'-s)(H_o- E)} B(r')\
\longrightarrow \ 0
\end{equation}
for any differentiable family of operators $B(r)$ which
does not depend on $\tau$.

Now consider the expression for $A^n_\tau$ obtained by inserting
$\1 = \bar P_E + P_E$ between the two right most factors of
$K_\tau$ in the integral which appears in eq.~\eqref{eq:intforA}.
Proceed with the term obtained from $P_E$ by inserting $\bar P_E +
P_E$ between the next two factors of $K$. Continue from right to
left in this way, expanding only the terms obtained from $P_E$. We
obtain an expression for $\bar P_E A^n_\tau(t,s) P_E $ as of sum
of $n$ terms, the $j$th term being
\begin{multline}
(-\im)^j \idotsint\displaylimits _{s \le r_n \le \ldots \le r_1
\le t} \di r_1 \ldots \di r_n \bar P_E K_\tau(r_1,s) \ldots
K_\tau(r_{n-j},s) \\ \times \bar P_E \e^{\im \tau
(r_{n-j+1}-s)(H_o- E)} \Lambda(r_{n-j+1}) P_E  \ldots \Lambda(r_n)
P_E \; ,
\end{multline}
which uniformly converges to zero by virtue of
eq.~\eqref{eq:barPEB}.  Since $A^n_\tau$ is a finite linear
combination of terms which uniformly tend to zero it does so as
well.
\end{proof}

It remains to show that $\norm{P \Omega_\tau(t,s) (\bar P -
P_\Delta)}$ is bounded by $1/\tau \Delta$.
\begin{lemma}\label{O}
Let $ P_1:= \chi(H_o \le E_1)$ and $P_2 := \chi(H_o \ge E_2)$ with
$E_2 > E_1$. Then
\begin{equation}\label{eq:Omegabound}
\Vert P_1 \Omega_\tau(t,s) P_2 \Vert \ \le \ \frac{C}{\Delta\tau}
\; ,
\end{equation}
where  $\Delta = E_2 - E_1$ and $C$ is a constant which does not
depend on $E_1$ or $E_2$.   The same inequality holds with $P_1, \
P_2$ interchanged.
\end{lemma}
\begin{proof}
As in the proof of Lemma~\ref{lem:PE} the idea is to prove a bound
on each term $A_\tau^n$ in the expansion for $\Omega_\tau$.  In
this case, we will show that
\begin{equation}\label{eq:Abound}
\norm{P_1 A_\tau^n(s,t) P_2} \ \le \ \frac{n}{\tau \Delta}
\frac{\alpha^n}{ (n-1)!}
\end{equation}
where $\alpha$ is a constant independent of $s,t$.  Summing these
bounds clearly implies eq.~\eqref{eq:Omegabound} -- see
eq.~\eqref{eq:series}.

The main step is to show that
\begin{equation}\label{eq:Kbound}
\Vert P_1 \int_t^sdrK_\tau(r,s)\ P_2 \Vert \ \le \
\frac{C}{\Delta\tau} \; ,
\end{equation}
and the same with $P_1$ and $P_2$ interchanged.   The idea is
that, since $K_\tau(r,s) = \e^{\im \tau(r-s)H_o} \Lambda(r)
\e^{\im \tau(s-t) H_o}$ and the spectral supports of $P_1$ and
$P_2$ are distance $\Delta$ apart, the integral over $r$ has a
highly oscillating phase of order $\tau \Delta$. For a rigorous
argument, however, it is convenient to use a commutator equation
and integration by parts to extract eq~\eqref{eq:Kbound}. This
method goes back to Kato \cite{k}.

The commutator $[H_o, X]$ might be ill defined if $H_o$ is
unbounded.  Thus we introduce a cutoff and work instead with
$[H_o, P_M X P_M]$ where $P_M = \chi(-M < H_o < M)$ and $M \in (0
, \infty)$.  At the end of the argument we take $M \rightarrow
\infty$.  The $X$ we have in mind is
\begin{equation}
X(r)\ := \ \frac{1}{2 \pi \im } \int_{\Gamma} \di z \, P_1 \, R(z)
\Lambda(r) R(z ) \, P_2 \; .
\end{equation}
where $R(z):= (H_0-z)^{-1}$ and the contour $\Gamma$ is the line
$\{E' + \im \eta \, : \, \eta \in \R\}$ with $E' = (E_2 + E_1)/2$.
A simple calculation yields
\begin{equation}
[H_o, P_M X(r) P_M]\ = \  P_M P_1 \Lambda(r) P_2 P_M\; .
\end{equation}
Therefore
\begin{equation}\label{eq:PKPX}
\begin{split}
P_M P_1 K_\tau(r,s) P_2 P_M \ =& \ [H_o, \e^{\im\tau(r-s)H_o} P_M
X(r) P_M \e^{\im
  \tau(s-r)H_o} ] \\ =& \ \frac{1}{\im \tau} \left ( \frac{\di}{\di r}
  ( P_M \e^{\im\tau(r-s)H_o} X(r) \e^{\im \tau(s-r)H_o} P_M) \right . \\ & \qquad \left .
  \phantom{\frac{\di}{\di r}}- \
   P_M \e^{\im\tau(r-s)H_o} \dot X(r) \e^{\im \tau(s-r)H_o} P_M\right
   ) 
   \; .
\end{split}
\end{equation}
However $X(r)$ and $\dot X(r)$ are uniformly bounded, $
\norm{X(r)} , \| \dot X(r)\|  \le  C /\Delta $,  so integrating
\eqref{eq:PKPX} yields
\begin{equation}
\Vert P_{M} \, P_1 \int_t^sdrK_\tau(r,s)\ P_2 \, P_{M} \Vert \ \le \
\frac{C}{\Delta\tau} \; .
\end{equation}
In the limit $M \rightarrow \infty$ this implies
eq.~\eqref{eq:Kbound} by lower semi-continuity of the norm.  
The second case with $P_1$ and $P_2$
interchanged follows with an obvious modification of $X$.

The rest of the argument is similar to the proof of
Lemma~\ref{lem:PE}.  We insert a decomposition of the identity $\1
= Q + \bar Q$ between the factors of $K$ in the integral
expression for $A_\tau^n$, eq.~\eqref{eq:intforA}. To apply
\eqref{eq:Kbound}, we should maintain a spectral gap between the
projections which sits to the left and right of $K$. Therefore we
define $Q_{j}:=\chi(H_{0}\le Q_{1}+j/n \Delta)$ for $j=0, ... , n$
and insert $1=Q_{j}+ \bar Q_{j}$ between the $j$th and $(j+1)$th
factors of $K$.   With these insertions, $P_1A_\tau^nP_2$ breaks
into $2^{n}$ terms and each term includes at least one factor of
the type $Q_{j} K_\tau(r_{j+1},s) \bar Q_{j+1}$ or $\bar Q_{j}
K_\tau(r_{j+1},s) Q_{j+1}$ where there is a gap of size $\Delta / n$
between the spectral supports of the two projections.  We apply
integration by parts to the integral over $r_{j+1}$ to obtain a
factor which may be bounded by eq.~\eqref{eq:Kbound}:
\begin{multline}
\int_0^{r_j} \di r_{j+1} \, Q_{j} K_\tau(r_{j+1},s) \bar Q_{j+1}
B(r_{j+1}) \\ = \ \int_0^{r_j} \di r_{j + 1} \, Q_j
\int_{r'}^{r_j} \di r' \, K_\tau(r',s) \bar Q_{j+1} \dot
B(r_{j+1}) \; .
\end{multline}
Elementary norm estimates and eq~\eqref{eq:Kbound} now show that
each of the $2^n$ terms is bounded by $n \beta^{n}/(\Delta\tau
(n-1)!)$ for some $\beta$ which implies
eq.~\eqref{eq:Abound} with $\alpha = 2 \beta$.
\end{proof}

\section{Integration by parts and the proof of lemma~\ref{lem:ibp}}\label{sec:BV}

Turning to the proof of Lemma~\ref{lem:ibp}, we note that the
spectral theorem provides the representation
\begin{equation}\label{eq:spectralthm}
f(H_o) \ = \ \int f(E) \di P_o(E)
\end{equation}
valid for bounded measurable $f$.  The goal is to integrate this
expression by parts thereby obtaining an expression involving $\di
f$ and $P_o(E) = \chi(H_o \le E)$.  This argument works precisely
when $f \in BV$ as we shall now explain.

The projection valued measure $\di P_o(E)$ is the differential of
$P_o(E) = \chi(H_o \le E)$ which is of bounded variation {\em in
the strong operator topology}. That is, for any $\psi \in \Hi$,
\begin{equation}\sup_{n \ge 1} \ \sup_{E_0 < \cdots < E_n
\in \R} \ \sum_{j=1}^n \|P_o(E_j) \psi - P_o(E_{j-1}) \psi \| \ <
\ \infty \; .
\end{equation}
We could equally well work with $P_o(E) = \chi(H_o < E)$ or a
number of other choices -- the distinction being meaningful only
if $H_o$ has point spectrum.

Since the function $P_o$ is SOT-continuous from the left at every
$E$, {\it i.e.}  $P_o(E-0) = P_o(E)$,  we may integrate
\eqref{eq:spectralthm} by parts whenever $f \in BV$ and everywhere
continuous from the right:\footnote{The extension of integration
by parts to functions in $BV$ is a standard part of real analysis
-- we direct the reader to \cite[Ch. 3]{folland} for details.}
\begin{multline}
f(H_o) \ = \ f(\infty) \1 - \int \di f(E) P_o(E) \; , \\ f \in BV
\mbox{ and continuous from the right.}
\end{multline}
For general $f \in BV$ this formula is replaced by
\begin{equation}\label{eq:integratedbyparts}
\begin{split}
f(H_o) \ = \ f(\infty) \1 \ -& \ \int \di f(E) \, \chi(H_o \le E)
\\ +& \ \sum_{E \in \R}
(f(E) - f(E+0)) \, \chi(H_o = E) \; .
\end{split}
\end{equation}
Note that $ \sum_{E \in \R} |f(E) - f(E+0)| \le  {\rm Var}(f)  < \infty$.
In particular, there can be only countably many $E \in \R$ for
which $f(E) \neq f(E+0)$.

Now suppose that
\begin{equation}\label{eq:Alim}
\sotlim \,  W_\tau( s) \, A \, W_\tau(s)^\dag \ = \ A
\end{equation}
uniformly in $s$ whenever $A = \chi(H_o \le E)$  or $A = \chi(H_o
\ge E)$ with $E \in \R$. Since
\begin{equation}
\chi(H_o = E) \ = \ \chi(H_o \le E) + \chi(H_o \ge E) - \1 \; ,
\end{equation}
eq.~\eqref{eq:Alim} also holds with $A = \chi(H_o = E)$. Now given
$f \in BV$, use \eqref{eq:integratedbyparts} to express $f(H_o)$
and find that
\begin{equation}
\sotlim \, W_\tau(s) f(H_o) W_\tau(s)^\dag \ = \ f(H_o)
\end{equation}
uniformly in $s$ by dominated convergence.

\section{Why the norm topology is inadequate -- an example}
\label{sec:example} The following example, due to Michael
Aizenman, is motivated by the consideration of systems with
dense point spectrum. 

We begin with a comment which, although mathematically trivial, already
contains a key observation.  If a family of operators, $A_\tau$,
converges (as $\tau \rightarrow \infty$) in the strong operator topology then 
although for any vector $\psi$ the family of vectors $A_\tau \psi$ is
convergent -- this is the very definition of SOT convergence --  nothing 
can be said regarding the {\em rate} of convergence.  In fact the essential difference
with norm convergence is that in the norm case  the vectors $A_\tau \psi$ all
converge at the same rate.

For a specific example, consider the countable collection of non interacting
two-level systems each perturbed by 
a weak perturbation (strength $\sim 1/\tau$):
\begin{equation}
H_{\tau} \ = \ H_o + \frac{1}{\tau} 
\Lambda \; , \qquad  H_o \ = \ \sum_{k=0}^\infty\oplus \; 
m_k \sigma_z \; , 
\qquad \Lambda \ = \ \sum_{k=0}^\infty\oplus \; \sigma_x\, ,
\end{equation}
with $\sigma_z, \sigma_x$ the Pauli spin matrices and $m_k = 1/k$.
The perturbation $\Lambda$ is time independent, 
but still Theorem~\ref{at} applies.  Of course, 
the unitary evolution $U_{\tau}$
associated with Hamiltonian $H_{\tau}$ decomposes
into a direct sum of two by two matrices $U^k_\tau$ each generated
by $H_{k,\tau} = (1/k) \sigma_z + (1/\tau) \sigma_x$ .

Let us choose $f(H_o) = P_o$ to be the spectral projection onto
negative energies: $P_o =\chi(H_o < 0)$. We will show, 
\begin{equation}
\limsup_{\tau \rightarrow \infty} \norm{U_\tau(\tau s)P_o
U^\dagger_\tau( \tau s)  \ - \ P_o} \ \ge \ \alpha(s)
\end{equation}
where $\alpha(s) > 0$ for every $s \in (0,1)$,
although, in accordance with Theorem~\ref{at},
\begin{equation}
\sotlim U_\tau(\tau s)P_o U^\dagger_\tau(\tau s) \ = \ P_o \; .
\end{equation}

Indeed, consider the particular sequence $\tau_n := n$.
For each $n$, the evolution of the two-level system
with $m_n:=1/n$ obeys
\begin{equation} \im \dot U^n_{\tau_n}(t) 
  \ = \ \frac{1}{n} (\sigma_z + \sigma_x)U^n_{\tau_n}(t)\, ,
\end{equation}
Thus, the matrix $V(s) := U^n_{\tau_n}(\tau_n s)$ is independent of 
$n$ and may be obtained by integrating
\begin{equation}
\im  \dot V(s) \ = \ (\sigma_z + \sigma_x) V(s)
\end{equation}
with initial condition $V(0) \ = \begin{pmatrix} 1 & 0 \\ 0 & 1
\end{pmatrix}$. It is now a simple matter to check that
\begin{equation}\label{eq:VsigV}
\norm{ V(s) \ \begin{pmatrix} 0 & 0 \\ 0 & 1
\end{pmatrix}  V(s)^\dag - \begin{pmatrix} 0 & 0 \\ 0 & 1
\end{pmatrix} }  \ >\ 0\,,
\end{equation}
for all $s\in(0,1)$.
Since, as mentioned above, $U_\tau$ is the direct
sum of the two by two matrices $U^k_\tau$ we have
\begin{equation}
\norm{U_{\tau_n}(\tau_n s) P_o U^\dag_{\tau_n}(\tau_n s)) - P_o}
\ \ge \ \norm{V(s) \begin{pmatrix} 0 & 0 \\ 0 & 1
\end{pmatrix} V(s)^\dag -
\begin{pmatrix} 0 & 0 \\ 0 & 1
\end{pmatrix}} \ > \ 0 \; .
\end{equation}

\section{The Schr\"odinger picture -- a theorem and a
  counter-example}\label{sec:conjecture} Theorem~\ref{at} describes the 
adiabatic limit of the {\em Heisenberg picture} of quantum
dynamics.  As for the {\em Schr\"odinger picture},  there is no 
reason to expect $U_\tau(\tau s)$ to converge to anything at all, since
even the unperturbed evolution, $\e^{-\im \tau s H_o}$, does not have
a large $\tau$ limit. 

With this in mind, it is natural to ask whether, in some sense,
$U_\tau$ is asymptotically equal to $\e^{-\im \tau s H_o}$.  
To test this idea we consider the evolution 
\begin{equation}
\Omega_\tau(s) \ := \ \e^{\im \tau s H_o} U_\tau(\tau s)
\end{equation}
which represents, physically, a process in which the system is evolved
forward in time according to the perturbed dynamics and then {\em backwards in
  time} according to the unperturbed dynamics.  
 
If $H_o$ admits an eigenfunction decomposition, i. e., if the spectrum of
$H_o$ is pure point, 
then a simple extension of the proof of Lemma~\ref{lem:PE} 
shows that $\Omega_\infty$ does exist and even allows us to calculate it.
\begin{thm}\label{Sat}
Let $H_o$ be a self-adjoint operator with only pure point spectrum.  
It $U_\tau$ satisfies the initial problem
\eqref{eq:IVP} with $H_\tau(t/\tau) = H_o + (1/\tau) \Lambda(t/\tau)$ where
$\Lambda(\cdot)$ is a self-adjoint family which is $L^1$ in norm, then
\begin{equation}
\sotlim_{\tau \rightarrow \infty} \, \e^{\im \tau s H_o}
U_\tau(\tau s) \ = \ \Omega_\infty(s) \;
\end{equation}
where $\Omega_\infty(s)$ is the unitary operator which commutes with $H_o$ and satisfies
the initial value problem
\begin{equation}\label{eq:omegainftyivp}
\begin{cases} \im \dot \Omega_\infty(s) \ = \ \left ( \sum_{E \in \sigma(H_o) } P_E
\Lambda(s) P_E \right ) \, \Omega_\infty(s)  \\ \Omega_{\infty}(0) \ = 
\ \1 \end{cases}\; ,
\end{equation}
where $P_E = \chi(H_o = E)$ is the orthogonal projection onto the space of
eigenvectors of $H_o$ with eigenvalue $E$.
\end{thm}
\noindent {\it Remarks:} 
\begin{list}{(\arabic{list})}
{\usecounter{list}  \setlength{\leftmargin=24pt}
\setlength{\labelsep=12pt} \setlength{\itemindent=12pt}}
\item When there is a uniform lower bound on the spacing between
neighboring eigenvalues, 
a classical result of Born and Fock \cite{BF} shows that the 
convergence occurs in the norm topology.
\item This theorem is of particular interest if $H_o$ has only 
dense point spectrum as is true of discrete random Schr\"odinger
operators in the large disorder regime -- see \cite{AM} for one 
perspective on this subject.
\end{list}
\begin{proof}
In the notation of Sec.~\ref{sec:proof}, the evolution
considered here -- $\Omega_\tau(s)$ -- is equal to the propagator 
$\Omega_\tau(s,0)$.  Thus, 
by lemma~\ref{lem:PE}, we see that $(\1 - P_E) \Omega_\tau(s)P_E$ converges to
zero uniformly in the strong operator topology for every $E$.  
To complete the proof of Theorem~\ref{Sat}, we let 
$\Omega_{\infty}(s)$ denote the solution to the initial value problem
\eqref{eq:omegainftyivp} and show that 
\begin{equation}
\sotlim P_E \Omega_\tau(s) P_E \ = \ P_E \Omega_\infty(s) P_E \; ,
\end{equation}
for each $E \in \sigma(H_o)$.

As we saw in the proof of Lemma~\ref{lem:PE}, 
$\Omega_\tau(s)$ satisfies a Volterra-type equation:
\begin{equation}
  \Omega_\tau(s) \ = \ - \im \int_0^s \di r K_\tau(r) \Omega_\tau(r) \; ,
\end{equation}
with $  K_\tau(r)  =  - \im \e^{\im\tau r H_o} \Lambda(r) \e^{- \im
  \tau r H_o}$.
Inserting into this expression $\1 = P_E + (\1 - P_E)$ between $K_\tau$ and
$\Omega_\tau$ we obtain
\begin{multline}\label{eq:partialvolt}
 P_E \Omega_\tau(s) P_E \ = \ P_E - \im \int_0^s \di r P_E \Lambda(r) P_E
 \Omega_\tau(r) P_E \\ - \ \im \int_0^s \di r P_E K_\tau(r) (\1 - P_E)
 \Omega_\tau(r) P_E \; ,
\end{multline}
since $P_E K_\tau(r) P_E = \Lambda(r)$.
The last term on the right side converges to zero uniformly in the
strong operator topology, again by Lemma~\ref{lem:PE}.  

It is clear from eq.~\eqref{eq:partialvolt} that if 
the limit of $P_E \Omega_\tau(s) P_E$ exists, 
then it obeys  the evolution equation \eqref{eq:omegainftyivp}.
However, eq.~\eqref{eq:partialvolt} does not
directly imply that the limit exists.  
On the other hand, $\Omega_{\infty}$ also satisfies a Volterra-type 
equation which when subtracted from \eqref{eq:partialvolt} yields,
for the difference ${\mathcal E}_\tau(s) = 
P_E\Omega_\tau(s)P_E -  P_E \Omega_\infty(s) P_E$,
\begin{equation}\label{eq:errorvolt}
{\mathcal E}_\tau(s) \ = \ -\im \int_0^s \di r P_E \Lambda(r) P_E
 {\mathcal E}_\tau(r)  \ + \ R_\tau(s) \; ,
\end{equation}
where the remainder $R_\tau(s)$ is the last term of \eqref{eq:partialvolt}
and converges to zero uniformly in the 
strong operator topology.  Using Gronwall's lemma \cite{Gr19}, we 
conclude from \eqref{eq:errorvolt} that
${\mathcal E}_\tau(s)$ converges to zero uniformly in the strong 
operator topology.

More concretely, let $\psi$ be any vector.  Then \eqref{eq:errorvolt} 
yields
\begin{equation}
\norm{ {\mathcal E}_\tau(s) \psi} \ \le \ \int_0^s \di r \norm{P_E 
\Lambda(r) P_E} \, \norm{{\mathcal E}_\tau(r) \psi}  \ + \ 
\norm{R_\tau(s) \psi} \; .
\end{equation}
>From this together with the classical Gronwall lemma we learn that
\begin{equation}
\norm{ {\mathcal E}_\tau(s) \psi} \ \le \ \left [ \sup_s \norm{R_\tau(s) 
\psi} \right ] \, \e^{\int_0^1 \norm{\Lambda(r)} \di r} \; .
\end{equation}
Since the factor in brackets converges to zero as $\tau \rightarrow 
\infty$ so does the right hand side of the above inequality.
\end{proof}

We conclude with an example which shows that in general $\Omega_\tau(s)$ need
not have a limit. Take  $H_o$ to be differentiation, $\im \,\di/\di x$,
 on $L^2(\R)$ and let
$\Lambda(s)$ be the operator of multiplication by a function
$\Lambda(x,s)$. Since $\e^{- \im t H_o}$ is a shift by $t$, the generator of
$K_\tau(s)$ is the operator of multiplication by $-\im \Lambda(x - \tau s,
s)$. Thus, since $K_\tau(s)$ is a commuting family,
\begin{equation}
  \Omega_\tau(s) \ = \ \e^{-\im \int_0^s \di r \Lambda(x - \tau r, r)} \; .
\end{equation}
If, for instance, $\Lambda(x,r)$ is the indicator function of the set $\cup_n
[2^{2n}, 2^{2n+1}]$ then the right hand side has no limit as $\tau \rightarrow
\infty$. 

In light of this example, it is interesting to ask what conditions may be
placed on $\Lambda(r)$ to ensure the convergence of $\Omega_\tau(s)$.  This
question is similar to the problem of existence of the wave operators 
which arises in scattering theory, and can be dealt accordingly.

\subsection*{Acknowledgments} 
We are grateful to M. Aizenman for the example presented in
Section~\ref{sec:example} as well as the
suggestion to consider the strong operator topology.
We are also indebted to Y. Avron for the insight that the norm adiabatic
theorem is linked with the presence of an intrinsic time scale.
This work was partially supported by
the NSF Grant PHY-9971149 (AE).

\vfill
\pagebreak

\end{document}